\documentclass[12pt,preprint]{aastex}

\input epsf.sty

\shorttitle{Very short gamma-ray bursts}
\shortauthors{Cline et al.}

\begin{document}
\bibliographystyle{plain}

\title{Study of Very Short GRB: New Results from BATSE and KONUS.
}

\author{D.B. Cline\altaffilmark{1}, B. Czerny\altaffilmark{2}, 
 C. Matthey\altaffilmark{1}, A. Janiuk\altaffilmark{2}, 
 S. Otwinowski\altaffilmark{1}}
\altaffiltext{1}{University of California Los Angeles, Department of Physics and 
Astronomy, Box 951447, Los Angeles, CA 90095-1547, USA; e-mail: stanislaw.otwinowski@cern.ch}
\altaffiltext{2}{Nicolaus Copernicus Astronomical Center, Bartycka 18,
            00-716 Warsaw, Poland}

\begin{abstract}
 Our recent studies of Very Short Gamma Ray Bursts (VSB) of duration $T_{90} \le 100$ ms
have indicated a significant angular asymmetry and a uniform $V/V_{max}$ distribution from
 the BATSE data.
Here we update these studies, and we extend our research to events observed with KONUS 
satellite which gives a new insight into the spectra not possible with the BATSE data.
KONUS has observed 18 events with $T_{90} \le 100$ ms duration.
These events display considerable numbers of photons with energies 
above 1 MeV and have photons above 5 MeV in all cases. These appear to
be some of the most energetic photons observed from any classes of GRB
to date.
\end{abstract}


\section{INTRODUCTION}

The  Long Bursts ($T_{90} > 2$ s) are known to originate at 
cosmological distances
with many identified counterparts
and they are widely believed to be related to the collapse of massive stars
\citep{vanpar97,metz97,fish92}.
Considerable information about the short bursts can be found 
\citep{ghir04,nakar02,maglio03,fred03,ram00}
including the very interesting
possibility that the long and short bursts come from the same
sources \citep{ghir04}. SWIFT results seem to indicate the merger scenario
\citep{berger05} but
further study of short bursts is necessary.
In this letter we discuss the properties of the Very Short Bursts ($T_{90} \le 100$ ms; 
hereafter VSB).  
We have found some unusual aspects of these events including:

   a) Very hard photon spectra compared to longer duration GRBs.

   b) A significant angular asymmetry on the Galactic Plane.

   c) A $<V/V_{max}>$ value consistent with 1/2.

In this paper we update the results of \cite{cline99,cline00,cline03} 
including new events. 
We also analyze the VSB observed with the KONUS detector
that have been recently reported \citep{konuscat}.
Sensitivity of KONUS detector above 1 MeV allows to draw firm conclusions 
on the
hard part of the VSB spectra.
Data analysis is given in Section 2 (BATSE results) and Section 3 (KONUS results), and the results are discussed in Section 4.

\section{FURTHER STUDY OF THE BATSE DATA}

 We use the data from the BATSE Current Gamma-Ray Burst
Catalog
\citep{batsecurrent}.
It contains the following data base:
 Basic (2702 events - last update Sep. 9, 2000),
 Duration (2041 events - last update Aug.17, 2000),
 CMAX/CMIN (1318 events - last update Aug. 1, 2003).
 The subsample of Short Bursts (hereafter SB), 
with $0.1$ s $ \le T_{90} < 2$ s , consists
of 448 bursts. There  are 51 bursts that have $T_{90} \le 0.1 $ s, and they
 form our subsample of VSB. 
We note that there has been evidence for an anisotropy in the SB
from a previous study \citep{maglio03}.  

\subsection{Anisotropic distribution of the VSB}

The angular distribution of VSB and SB in Galactic Coordinates is shown in 
Fig. 1. The sky was divided
into 8 equal regions.
 In the case of isotropic distribution the number of bursts 
in each region should
be described by the Poisson distribution. The histograms of such values are
shown in Fig. 1. The probability
of any number of bursts in a single zone multiplied by the number of zones (8)
 is shown with full circles.
For 448 SB ($0.1$ s $\le T_{90} < 2$ s)  we see conformity with 
isotropic distribution, but for VSB ($T_{90} \le 0.1$ s)
the observed distribution is strongly improbable for isotropic distribution.
The number of bursts in one of the regions (roughly in the direction of the 
Galactic Anticenter) is 20, which is much higher than the expected average 
of 51/8.
 The probability to find twenty or more
 events (from the total number of fifty one) in the region of 1/8 area is 0.00007.
This result argues for other explanation than the statistical fluctuation.
Background in the direction of the Galactic Center is $12 500 \pm 1000$ 
cts~s$^{-1}$,
while the mean level of the background outside the region is 
$13 800 \pm 1300$ cts~s$^{-1}$, and the total number of SB in this regions is slightly lower than the expected average (but within 
the expected error). Therefore, background anisotropy cannot
be responsible for the observed distribution of the VSB across the sky.

We support the choice of $T_{90}$ = 0.1 s to discriminate VSB from SB 
with the following consideration. 
In Fig.2 we show the distribution of the excess number
 of the GRBs as
a  function of $T_{90}$. The excess from the isotropic distribution is 
calculated as a difference between
the GRBs number in the chosen region 
 and the sum of all GRBs 
number in other
regions divided by seven. This study shows that
the incompatibility with isotropic distribution is seen only for GRBs with
$T_{90} \le 0.1$ s. We observe the total excess of $15.71 \pm 4.52$ bursts 
in this region. 

\subsection{\textbf{$<V/V_{max}>$} test of radial distribution}

We have also reanalyzed the overall radial distribution of the VSB and  
the SB using the standard $<V/V_{max}>$ test
\citep{schmdt88}.
We use the CMAX/CMIN table from BATSE catalog as an input for $V/V_{max}$ 
calculations:
          $V/V_{max}$ = $(C/C_{min})^{-3/2}$
and we follow the standard algorithm given in \cite{batsecurrent}. Eight 
events in 
the table CMAX/CMIN have all values lower than the accepted trigger.
We can either reject these events, or attribute them the value 
close to the detection threshold
($V/V_{\rm max}$ $\approx$ 0.9). 
The choice does not affect significantly  the measured value of $V/V_{max}$
(see Table~\ref{tab0}).

The result of $V/V_{max}$ distributions and mean values are presented
in Fig. 3. SB appear to come from cosmological distances
given the $V/V_{max}$ distribution we see  
 since this is 
similar to the distribution for the Long GRBs.
No cosmological effects are seen in the
distribution of VSB.  

\section{STUDY OF THE VSB FROM THE KONUS SAMPLE}

   In Table~\ref{tab1} are listed the candidates for VSB from
 KONUS - 18 events!
This is approximately what would be expected from the relative KONUS 
and BATSE exposure and confirms that the BATSE events are real. In the case,
when the same event (the same time of registration) was observed in both
KONUS and BATSE experiments both $T_{90}$ values are shown in Table~\ref{tab1}
\citep{konuscat}.
Independent detections by satellites other than KONUS are also marked, including 
the trigger number  
(B-BATSE, U-Ulysses, N-NEAR). We accept as VSB events where no other significant
 maximum (above background) appears within 30 seconds after the trigger.

The KONUS detector has a significantly larger energy acceptance
out to tens of MeV.
The BATSE detector has a smaller NaI absorber and is not sensitive to
1-10 MeV photons. 
Therefore the KONUS detector offers the first chance to measure higher
energy photons from VSB.
However, there are no corresponding coordinates
for the KONUS events as in the case of BATSE. We therefore
use the KONUS data mainly for spectral information.

All VSB show appreciable number of photons above 1 MeV energy.
This follows the trend observed in the BATSE data that the VSB events
have a hard energy spectrum.
All events also show gamma rays above 5 MeV.
In Fig.4 we compare the mean energy of SB and 
VSB for KONUS events.
We observe that in the MeV region the spectra of VSB are significantly harder 
than the spectra of SB. The spectra start to be flatter above 3 MeV and
the effect in the case of VSB is again stronger.

\section{DISCUSSION}

This extended analysis of the BATSE data and the results of the KONUS
detector continue to suggest that the VSB events form a separate class 
of events from the SB. 
Strong anisotropy and $<V/V_{\rm max}>=0.5$ consistent with Euclidean space 
strongly argue 
for the local origin of this phenomenon. Spectra of VSB are very hard, 
supporting the 
claim that a different mechanism is here at work.

Detected counterparts of Long Bursts at other wavelengths established the
relation of those events to the massive star collapse; no such support was
available for SBs \citep{zhang04}. Therefore,
several possibilities for the origin of this class of bursts were 
proposed, e.g. \cite{cline03}.
This includes neutron star mergers \citep{good87,dar99} and neutron star collapse
to strange-quark stars \citep{dar99}. 
There were also suggestions that the nature of Long and Short Bursts is
the same \citep{yama04} although SB are on average harder 
\citep{kouv93,ghir04}, have $<V/V_{\rm max}>$ closer to 1 and show traces of 
anisotropy \cite{maglio03}.
VSBs are extreme examples being still harder and showing even more 
spectacular anisotropy, clearly indicating towards the local origin of
this class of events.

Primordial black hole evaporation, as an explanation of the existence of 
the VSB class, is also possible (for details, see \cite{cline2000}).
The energy expected from such 
phenomenon is of order of $M_*c^2$, and the currently evaporating black 
holes have 
masses $M_* \sim 10^{15}$ g 
\citep{carr,becken}.
Such energy, if converted with an efficiency 
$f_{\gamma}$ into the form of $\gamma$-rays 
collimated in a solid angle $\Omega$,  would give the observed fluence
$F = 8 \times 10^{-9}f_{\gamma}({M_* \over 10^{15}{\rm g}})
({1 {\rm kpc} \over D})^2 ({4 \pi \over \Omega}) ~~~ {\rm [erg~cm^{-2}]}$.
The weakest VSB (see Table~\ref{tab1}) can come from distances of $\sim 300$ pc if uncollimated,
and from further out if some collimation is involved. However, less standard
approach to primordial black hole mass evaporation gives much broader range of
numbers 
\citep{send05}.

We also note that there is some confusion between the Short GRB and the SGR
super flares \citep{hur05}. However, the very hard spectrum of the VSB 
would seem to separate those types of events.

Plausible detection of X-ray afterglows \citep{fred03}, and recent
claims of the identification
of a host galaxy for two SB Swift bursts \citep{bloom05,berger05} might
point toward merger scenario. One of the localized bursts (GRB~050724) 
is a typical
SB ($T_{90} \sim 0.26$ s). The other one (GRB~050509b) is intermediate between
a VSB and a SB. Its Galactic Coordinate is not in the excess region 
of Fig 1a and the limited energy acceptance of Swift does not allow
a test of the presence of MeV photons to compare with other VSB
studied here. More SWIFT data may clear the issue.

\section{CONCLUSIONS}

In this paper we have provided a new analysis of the most recent VSB
events from BATSE. We have also studied the VSB from KONUS for the
first time. We show that the GRBs with time duration below 100 ms 
appear to form a separate class of GRBs. The KONUS data show that these
events have a very hard photon spectrum up to 10 MeV - this effect was
already hinted in the BATSE hardness data. The events appear to originate
nearby within the Galaxy as the asymmetry plot (Fig. 1.a) and
the $V/V_{max}$ distribution (Fig. 2.a) indicate.
 This effect is statistically significant. In previous studies of this subject, we have suggested that such events could come from
Primordial Black Hole evaporation 
\citep{cline03}, with supplementary details in 
\cite{cline2000}. The other local sources such as neutron stars should also be 
candidates.

{\it Acknowledgments.}  
This work
was supported in part by grants 2P03D00322 and PBZ-KBN-054/P03/2001.  
We thank the referee for information on Short Burst studies.

\clearpage


\begin{figure}[!h]
\centering
\epsfxsize = 120 mm \epsfbox{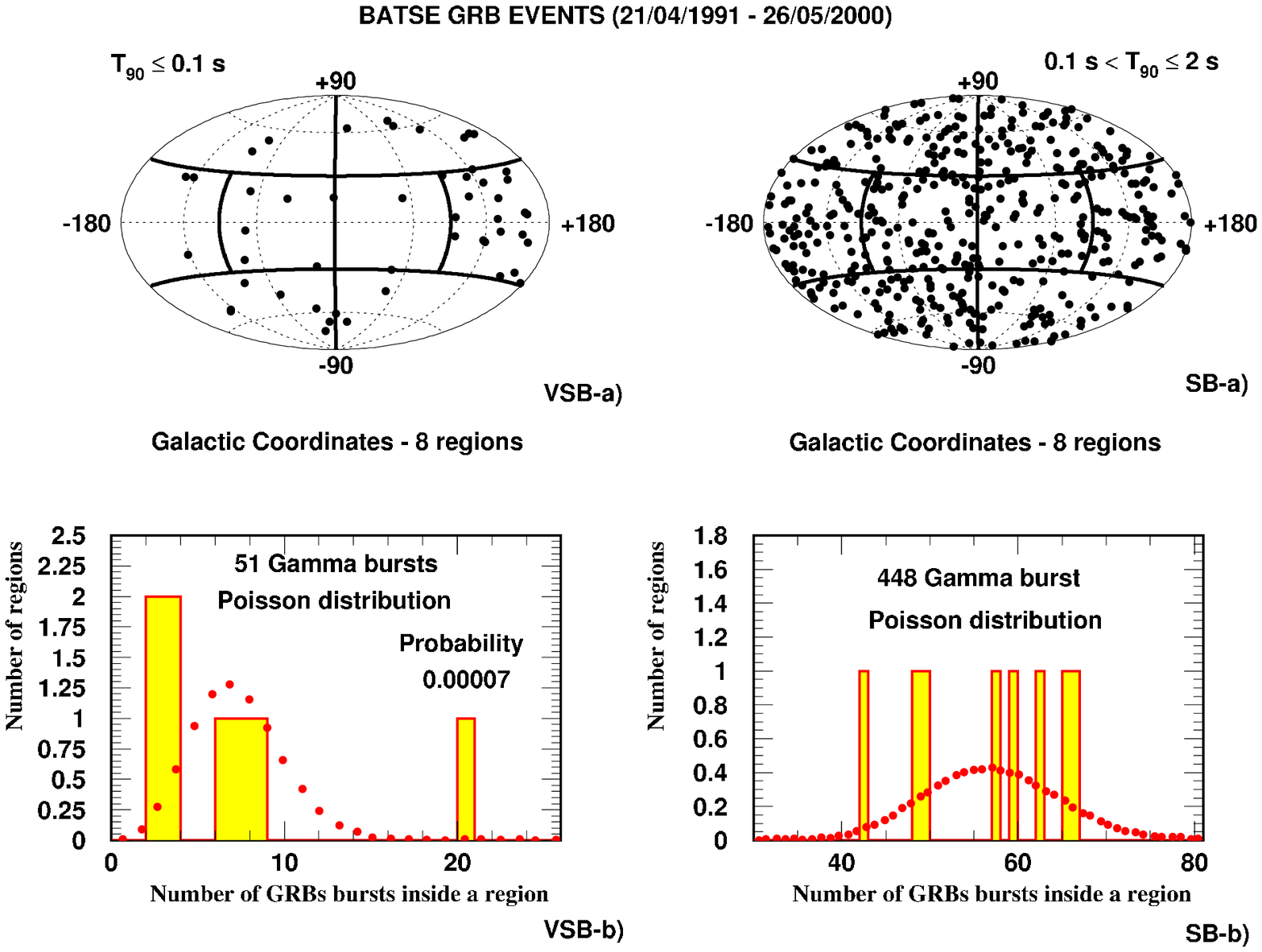}
\caption{Angular distribution of the GRBs in Galactic Coordinates 
  and the corresponding histograms, in comparison
 with Poisson distribution predictions for two different $T_{90}$ ranges
(full circles).}
\label{Fig.1}
\end{figure}

\clearpage


\begin{figure}[!h]
\centering
\epsfxsize = 130 mm \epsfbox{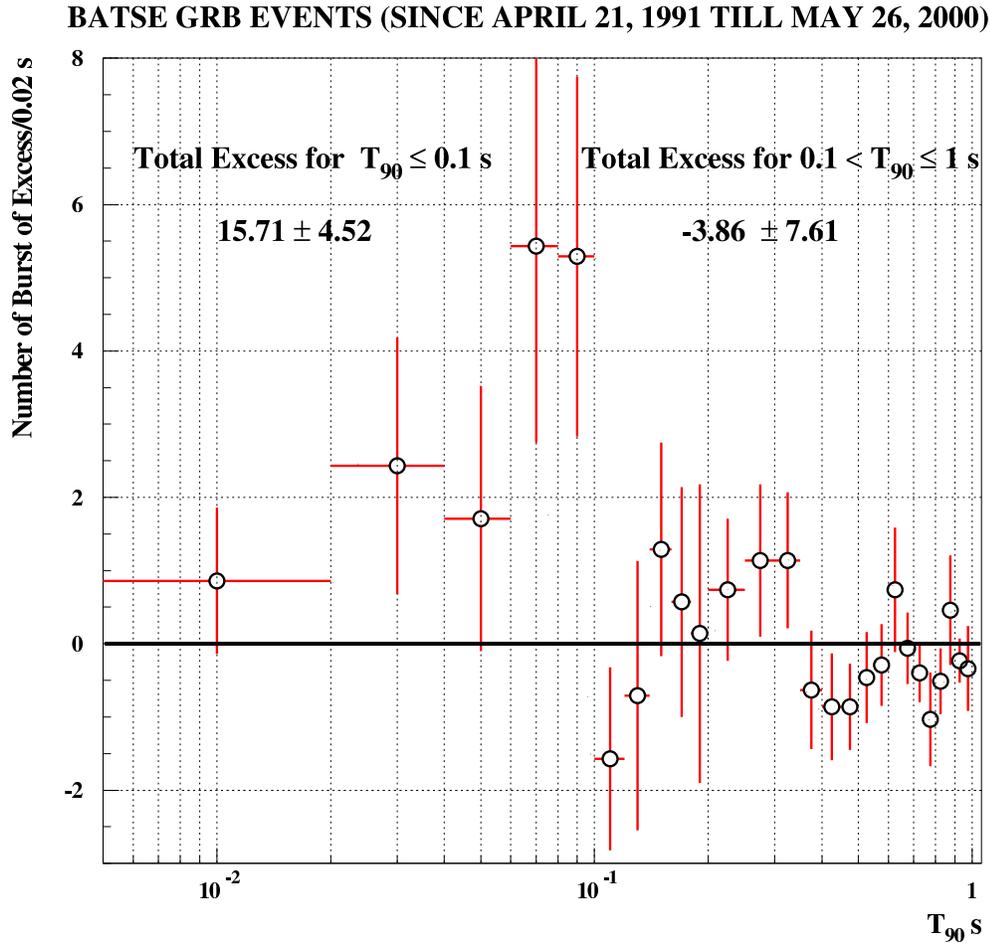}
\caption{The excess in the GRBs number inside the chosen region in the 
Galactic Plane (see Fig. 1)
as a function of $T_{90}$.}
\label{Fig.2}
\end{figure}

\clearpage

\begin{figure}[!h]
\centering
\epsfxsize = 150 mm \epsfbox{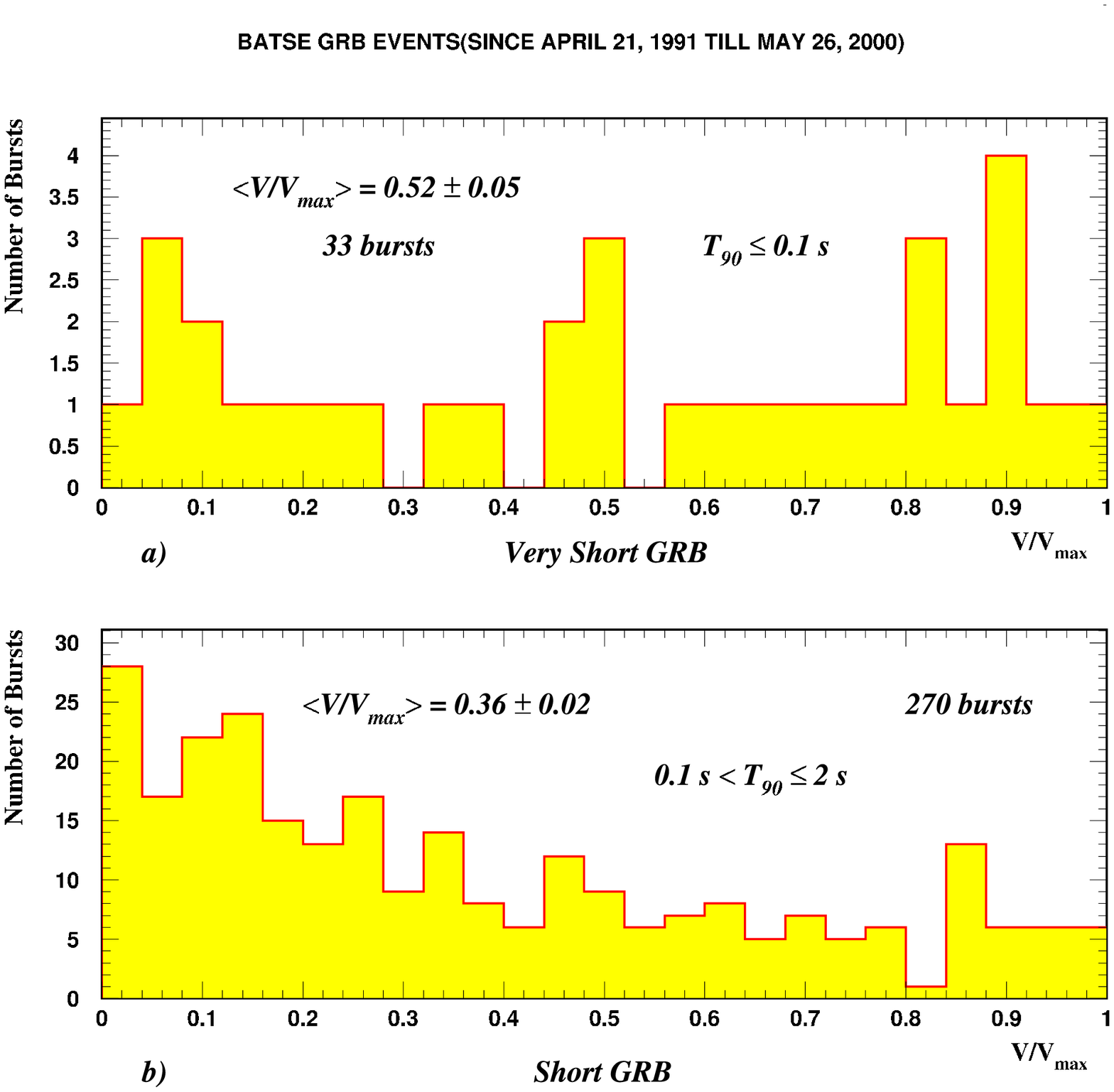}
\centering
\caption{Distribution of the $V/V_{max}$ for BATSE events.}
\label{Fig.3}
\end{figure}

\clearpage


\begin{figure}[!h]
\centering
\epsfxsize = 150 mm \epsfbox{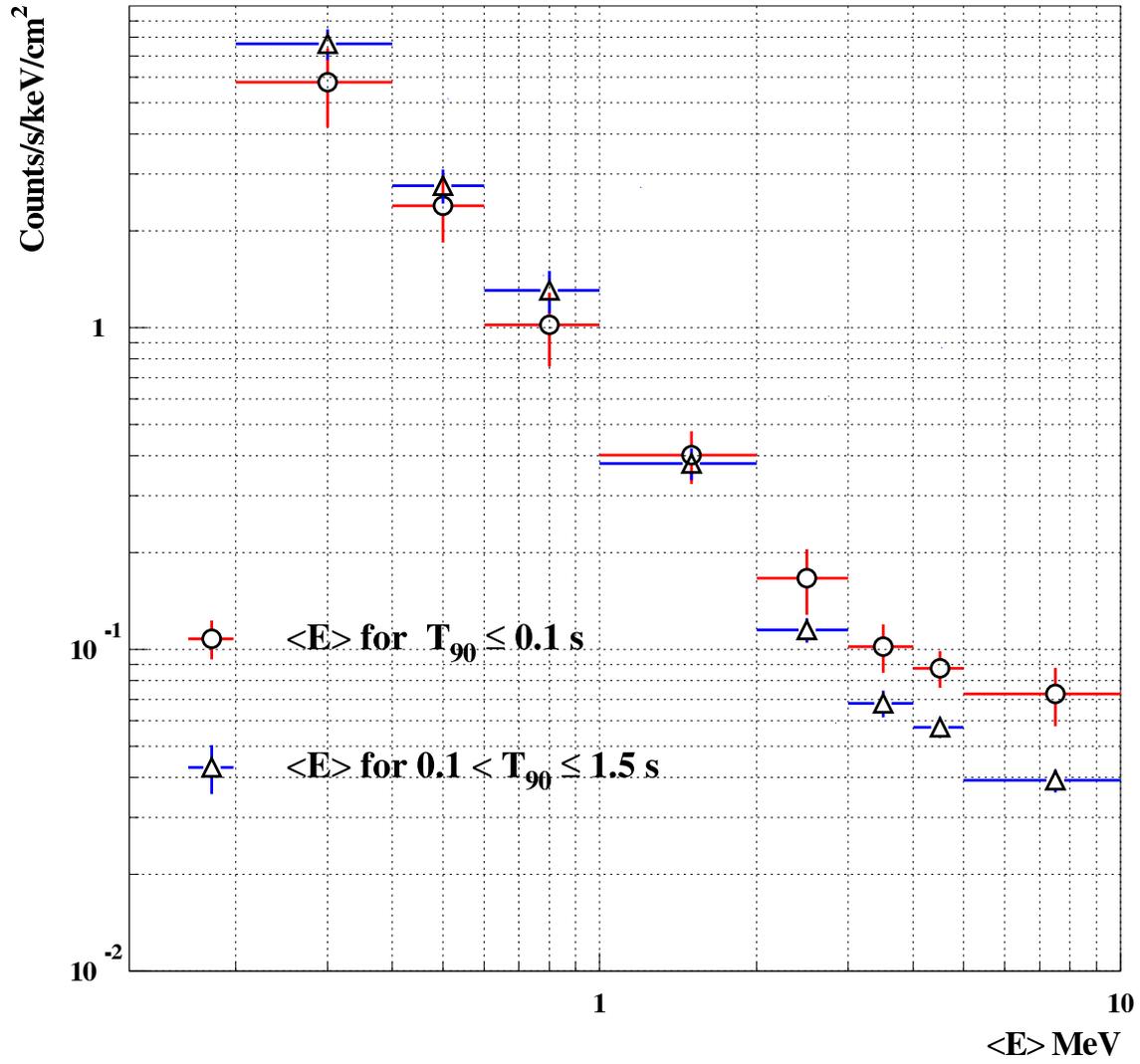}
\caption{Mean Energy distribution for KONUS Short and Very Short Bursts.}
\label{Fig.4}
\end{figure}

\clearpage

\begin{table}
\begin{center}
\caption{The influence of the uncertain BATSE detections on the measured $<V/V_{\rm max}>$
}
\label{tab0}

\begin{tabular}{rccc}
\tableline
\tableline
  $C/C_{\rm min}$ $<$ 1   &  VSB   &  SB  & $T_{90}$ $>$ 2 s \\
\tableline
  rejected events & 0.48 $\pm$ 0.05  & 0.37 $\pm$ 0.02 & 0.30 $\pm$ 0.01 \\
  accepted events & 0.52 $\pm$ 0.05  & 0.36 $\pm$ 0.02 & 0.30 $\pm$ 0.01 \\
\tableline
\end{tabular}
\end{center}
\end{table}

\clearpage

\begin{table}
\begin{center}
\caption{List of KONUS GRBs with $T_{90} \le 0.1$ s.}
\label{tab1}
\end{center}
\begin{center}
\begin{tabular}{rlcccccc}

\tableline
\tableline
NN & Burst  & $T_{90}$ [s] & Peak Flux &Fluence &Energy & Location& $T_{90}$ [s] \\
 &Name  & (KONUS)  & erg ${{\rm cm}^{-2}}$ ${{\rm s}^{-1}}$ &erg ${{\rm cm}^{-2}}$ &Interval keV & & (BATSE)\\

\tableline
110  & 010420a & 0.01 & 1.2e--005 & 2.3e--007 &15-1000 & -& -\\
 32  & 970625a & 0.03 & 3.8e--005 & 9.4e--007 &15-3000 & -& - \\
 18  & 960803  & 0.05 & 2.1e--005 & 8.9e--007 &15-5000 & B(tr.5561)& 0.10\\
 57  & 980904  & 0.05 & 3.5e--005 & 1.3e--006 &15-2000 & B(tr.7063)& 0.13\\
 27  & 970427  & 0.06 & 1.0e--005 & 3.9e--007 &15-1000 & B(tr.6211)& 0.07\\
 37  & 970921  & 0.06 & 1.0e--004 & 4.5e--006 &15-8000 & -&  -\\
 79  & 970427  & 0.06 & 1.0e--005 & 3.9e--007 &15-1000 & -&  -\\
 6   & 950610b & 0.07 & 1.6e--005 & 7.1e--007 &15-1000 & -&  -\\
 75  & 990504  & 0.08 & 1.5e--005 & 1.2e--006 &15-2000 & -&  -\\
 83  & 990831  & 0.08 & 5.9e--006 & 3.8e--007 &15-1000 & -&  -\\
 40  & 971118  & 0.08 & 2.7e--006 & 1.7e--007 &15-1000 & B(tr.6486)& 0.07\\
 86  & 991226b & 0.08 & 7.7e--006 & 3.9e--007 &15-1000 & -&  -\\
 117 & 020116  & 0.08 &      -    &     -     &   -    & -&  -\\
 34  & 970704  & 0.09 & 1.5e--003 & 4.2e--005 &15-10000 &B(tr.6293)& 0.19 \\
 95  & 000607  & 0.09 & 1.2e--004 & 5.4e--006 &15-3000 & U,N,GCN693& -\\
 119 & 020218a & 0.09 &      -    &      -    &  -     & -& -\\
 48  & 980330a & 0.10 & 5.2e--005 & 3.0e--006 &15-5000 & B(tr.6668)& 0.12\\
 112 & 010616  & 0.10 & 1.8e--005 & 1.5e--006 &15-1000 & -& -\\
\tableline
\end{tabular}
\end{center}
\end{table}

\end {document}